\documentclass[%
 superscriptaddress,
preprint,
 amsmath,amssymb,
 aps,
]{revtex4-2}

\usepackage{graphicx}% Include figure files
\usepackage{dcolumn}% Align table columns on decimal point
\usepackage{bm}% bold math
\usepackage{hyperref}% add hypertext capabilities
\begin{document}

% \preprint{APS/123-QED}

\title{Constraints on Einstein-dilaton Gauss-Bonnet gravity with Taiji}% Force line breaks with \\
% \thanks{A footnote to the article title}%

\author{Weilong Luo}
\email{luoweilong21@mails.ucas.ac.cn}
\affiliation{School of Fundamental Physics and Mathematical Sciences, Hangzhou Institute for Advanced Study, University of Chinese Academy of Sciences, Hangzhou 310024, China}
\affiliation{Institute of Theoretical Physics, Chinese Academy of Sciences, Beijing 100190, China}
\affiliation{University of Chinese Academy of Sciences (UCAS), Beijing 100049, China}
% \affiliation{xxxxxxxx}

\author{Chang Liu}
\email{liuchang@yzu.edu.cn}
\affiliation{Center for Gravitation and Cosmology, College of Physical Science and Technology, Yangzhou University, Yangzhou, 225009, China}
\affiliation{School of Fundamental Physics and Mathematical Sciences, Hangzhou Institute for Advanced Study, University of Chinese Academy of Sciences, Hangzhou 310024, China}
\affiliation{School of Physical Sciences, University of Chinese Academy of Sciences, No.19A Yuquan Road, Beijing 100049, China}

\author{Zong-Kuan Guo}
\email{guozk@itp.ac.cn}
\affiliation{CAS Key Laboratory of Theoretical Physics, Institute of Theoretical Physics, Chinese Academy of Sciences, Beijing 100190, China}
\affiliation{School of Physical Sciences, University of Chinese Academy of Sciences, No.19A Yuquan Road, Beijing 100049, China}
\affiliation{School of Fundamental Physics and Mathematical Sciences, Hangzhou Institute for Advanced Study, University of Chinese Academy of Sciences, Hangzhou 310024, China}

% \date{\today}% It is always \today, today,
%  but any date may be explicitly specified

\begin{abstract}

In the 2030s, space-based gravitational-wave (GW) detectors will exhibit unprecedented sensitivity in the millihertz frequency band, greatly expanding the potential for testing theories of gravity compared to ground-based GW detectors.
Inspired by effective string theory, Einstein-dilaton Gauss-Bonnet (EdGB) gravity introduces an extra dilaton scalar field that is directly coupled to higher curvature terms.
Here, we investigate the capability of Taiji to constrain the parameters of EdGB gravity by analyzing GWs from massive black hole binaries (MBHBs). 
We utilize the parameterized post-Einsteinian (ppE) waveform with the leading order EdGB corrections for the inspiral phase of MBHBs.
The constraints on the coupling constants are obtained by performing Fisher matrix analysis. 
With different mass ratios and spins $\chi_i$ at redshifts $z=2,3,4,5$, the $1\sigma$ bounds on the parameter $\alpha$ have the same order of magnitude: $\sqrt{\alpha}\sim 10^7$m.

\end{abstract}

%\keywords{Suggested keywords}%Use showkeys class option if keyword
%display desired
\maketitle

%\tableofcontents

\section{\label{sec:intro}Introduction}

General relativity (GR), the most successful theory of gravity today, has passed numerous tests~\cite{Will:2014kxa}.
In recent years, analyses and investigations of data obtained from the GW detectors have found no signs of contradictions with Einstein's theory~\cite{LIGOScientific:2019fpa,LIGOScientific:2020tif,LIGOScientific:2021sio}.

Space-based GW missions in the millihertz band have the potential to detect a large number of spatially distributed GW signals, some of which may have large signal-to-noise ratios (SNR) and some of which may last for long periods~\cite{Amaro-Seoane:2012aqc,Baker:2019nia,Hu:2017yoc}.
This makes GW detection in the millihertz band important for fundamental physics and cosmology~\cite{LISA:2022kgy,LISA:2022yao,LISACosmologyWorkingGroup:2022jok}.
Future space-based detectors such as LISA, Taiji, and Tianqin may be able to detect GW signals from MBHBs with a very high SNR, providing an excellent opportunity to investigate the nature of gravity.

Scalar Gauss-Bonnet gravity, extensively explored in the literature, is a gravitational theory that goes beyond GR~\cite{Nojiri:2005vv,Yagi:2012gp,Antoniou:2017hxj,Antoniou:2017acq,Nojiri:2010wj,Nojiri:2017ncd}.
In this theory, a dynamic scalar field is intricately coupled to a Gauss-Bonnet invariant, which is constructed from a specific combination of scalars with squared curvature. 
This coupling is governed by a constant $\alpha$ with units of squared length.
Various theories can be obtained depending on the specific type of coupling being considered.
These include the shift-symmetric theory (linear coupling)~\cite{Yagi:2011xp,Barausse:2015wia}, EdGB gravity~\cite{Kanti:1995vq,Torii:1996yi,Guo:2008hf,Maeda:2009uy,Herrero-Valea:2021dry}, which inspired from string theory and inflation~\cite{Odintsov:2020sqy,Oikonomou:2021kql,Odintsov:2020xji,Odintsov:2023weg,Odintsov:2023lbb}, as well as theories that allow for spontaneous scalarization of black holes and neutron stars (quadratic coupling serves as an example)~\cite{Doneva:2017bvd,Doneva:2017duq}. 
The propagation of GW in Einstein-Gauss-Bonnet gravity was discussed in Ref.~\cite{Nojiri:2023mbo,Nojiri:2023jtf,Elizalde:2023rds}

The anticipated bounds on EdGB gravity have been investigated with the spaced-based detectors, including LISA and Tanqin for inspiral signals~\cite{Gnocchi:2019jzp,Shi:2022qno}, as well as Taiji for ringdown signals~\cite{Shao:2023yjx}.
In this paper, we explore the capability of Taiji to constrain EdGB gravity using the inspiral signals from MBHBs.
By employing the Fisher matrix analysis, we determine the $90\%$ credible upper limit for $\sqrt{\alpha}$ with the $-1$PN correction in the GW phase. 
These bounds are found to be consistent across different spin configurations.

The paper is organized as follows. 
In Sec.~\ref{sec:edgbwave}, we introduce the ppE waveform for EdGB gravity. 
In Sec.~\ref{sec:method}, we outline the methodologies employed and key assumptions made in our calculations. 
We present results and summary in Sec.~\ref{sec:results} and \ref{sec:sum}.
Throughout this paper, we adopt the convention of setting the gravitational constant and speed of light equal to unity ($G=c=1$).

\section{\label{sec:edgbwave}The parametrized post-Einsteinian waveform in Einstein-dilaton Gauss-Bonnet gravity}

First, we introduce EdGB theory of gravity in the framework of scalar Gauss-Bonnet theory and explain its modification to the gravitational waveform in comparison to GR.
The action of scalar Gauss-Bonnet theory can be written as~\cite{Nojiri:2005vv,Yagi:2012gp,Antoniou:2017hxj,Antoniou:2017acq}
\begin{equation}
    S=\int d^4x\sqrt{-g}\left(\frac{R}{16\pi}-\frac{1}{2}\partial_\mu\varphi\partial^\mu\varphi+\alpha f(\varphi) \mathcal{R}_{GB}^2\right)+S_m \, ,
\end{equation}
where $S_m$ represents the matter action, $g$ denotes the determinant of the metric $g_{\mu\nu}$, $R$ is the Ricci scalar, and $\varphi$ denotes the `dilaton' scalar field.
In this paper, we focus on the EdGB gravity with $f(\varphi)=e^{\varphi}/4$~\cite{Kanti:1995vq,Moura:2006pz,Pani:2009wy,Ripley:2019irj}. 
The parameter $\alpha$ characterizes the coupling strength and has the dimensions of length squared, $\mathcal{R}^2_{GB}$ is the curvature-dependent Gauss-Bonnet invariant:
\begin{equation}
    \mathcal{R}^2_{GB}=R_{\mu\nu\rho\sigma}R^{\mu\nu\rho\sigma}-4R_{\mu\nu}R^{\mu\nu}+R^2 \, .
\end{equation}
To simplify the calculation, the small coupling approximation is adopted, where the contributions from EdGB theory are treated as small perturbations on GR~\cite{Mignemi:1992nt,Pani:2011gy}.
It is convenient to define a dimensionless coupling constant as
\begin{equation}
    \frac{16\pi \alpha^2}{\ell^4}\ll1 \, ,
\end{equation}
where $\ell$ represents the characteristic length scale of a given system~\cite{Yagi:2015oca,Lyu:2022gdr}.
The inequality is valid in the small-coupling regime.

In our analysis, we adopt the widely used ppE formalism\cite{Yunes:2009ke} to incorporate the EdGB corrections into the inspiral stage of the GW signals from MBHBs and neglect the merge-ringdown part of the signals for simplicity.
Although signals from the merger-ringdown phase will contribute considerable SNR, the duration is typically short.
Our results here will provide a conservative estimation of Taiji's capability to constrain EdGB gravity.
At $-1$PN order, the corrections only enter the phase of the waveform~\cite{Cutler:1994ys,Buonanno:2009zt,Barausse:2016eii}.  
In the frequency domain, the ppE waveform is given by~\cite{Yunes:2009ke}
\begin{equation}
    h_{ppE}(f)=h_{GR}(f)e^{i\beta u^b} \, .
\end{equation}
Here $u=\pi \mathcal{M}f$, $\mathcal{M}=M\eta^{3/5}$ is the chirp mass, $\eta=m_1m_2/M^2$ is the symmetric mass ratio, $m_1 $ and $m_2$ are the component masses of the binary, $M=m_1+m_2$ is the total mass.
For the GR part of the waveform $h_{GR}$, we use the IMRPhenomB waveform for non-precessing, spinning MBHBs~\cite{Ajith:2009bn}, with a Newtonian amplitude and averaged sky direction. 
EdGB gravity is described under the ppE framework by the following parameters\cite{Yunes:2013dva}:
\begin{equation}
    b=-\frac{7}{3} \, ,
\end{equation}
\begin{equation}
    \beta=-\frac{5(m_1^2s_2-m_2^2s_1)^2}{7168\eta^{18/5}M^4}\zeta \, ,
\end{equation}
where $\zeta=16\pi \alpha^{2}/M^4$.
Here $s_{i=1,2}$ are the factors related to the scalar charges of the black hole~\cite{Yunes:2016jcc,Berti:2018cxi}, which are given by
\begin{equation}
    s_i=2\frac{(\sqrt{1-\chi_i^2})-1+\chi_i^2}{\chi_i^2} \, .
\end{equation}
The normalized spin angular momentum~\cite{Yunes:2016jcc}
\begin{equation}
    \chi_i=(\vec{S}_i\cdot\hat{L}_i)/m_i^2  \, ,
\end{equation}
where $\vec{S}_i$ is the spin angular momentum, $\hat{L}_i$ is the  orbital angular momentum~\cite{Wang:2021jfc}.

\section{\label{sec:method}Method}
In this study, we use the Fisher matrix~\cite{Cutler:1994ys,Poisson:1995ef,Berti:2004bd,Yagi:2009zm} to calculate constraints on the ppE parameter $\beta$ and the EdGB coupling constant $\sqrt{\alpha}$.
The Fisher matrix method holds under the assumptions of large SNR and stationary Gaussian noise~\cite{Vallisneri:2007ev}. 
Here we assume that the instrument noise is Gaussian and stationary, as described by the theoretical model~\cite{Robson:2018ifk,Ruan:2018tsw}. 
The noise-weighted inner product between two functions, $A(t)$ and $B(t)$ is defined as 
\begin{equation}
    (A|B)=2 \int_{f_{low}}^{f_{high}}\frac{\tilde{A}(f)\tilde{B}^*(f)+\tilde{A}^*(f)\tilde{B}(f)}{S_n(f)}df \, ,
\end{equation}
where $\tilde{A}(f)$ and $\tilde{B}(f)$ are the Fourier transforms of $A(t)$ and $B(t)$, $S_n(f)$ is the sensitivity of Taiji.
The sky-averaged SNR $\rho$ of the GW $h$ can be expressed in terms of the inner product as
\begin{equation}
    \rho^2=(\tilde{h}|\tilde{h})=4\int_{f_{low}}^{f_{high}}\frac{|\tilde{h}(f)|^2}{S_n(f)}df \, .
\end{equation}
In our analysis, we set the innermost stable circular orbit frequency $f_{isco}$ of the system as the high-frequency cutoff $f_{high}$:
\begin{equation}
    f_{high}=f_{isco}=\frac{1}{6^{3/2}\pi M}   \, . 
\end{equation}
And the initial frequency is determined by the  observation time $T_{obs}$ as~\cite{Berti:2004bd}
\begin{equation}
    f_{low}=4.149\times10^{-5}\left(\frac{\mathcal{M}}{10^6M_\odot}\right)^{-5/8}\left(\frac{T_{obs}}{1\mathrm{yr}}\right)^{-3/8}  \,\mathrm{Hz} \,.
\end{equation}
The GW signals from inspiraling MBHBs can last for several months in mHz band~\cite{LISA:2017pwj}.
Here we assume that the signals will reach the frequency $f_{isco}$ after half a year of observations, which means $T_{obs} = 0.5 \mathrm{yr} $.
The sky-averaged sensitivity of space-based GW detectors can be written as~\cite{Robson:2018ifk} 
\begin{equation}
    S_n(f)=\frac{10}{3L^2}\left(P_{op}+2(1+\cos^2(f/f_*))\frac{P_{acc}}{(2\pi f)^4}\right)\left(1+\frac{6}{10}\left(\frac{f}{f_*}\right)^2\right) \, .
\end{equation}
For Taiji, the arm length $L=3\times10^9$m, $f_*=1/(2\pi L)$ is the transfer frequency. 
The budget of the acceleration noise~\cite{Luo:2019zal,Ruan:2018tsw}
\begin{equation}
    P_{acc}=(3\times10^{-15})^2 \left(1+\left(\frac{0.4 \mathrm{mHz}}{f}\right)^2\right) \mathrm{m}^2 \, \mathrm{s}^{-4} \, \mathrm{Hz}^{-1} \, .
\end{equation}
The optical path noise of Taiji is 
\begin{equation}
    P_{op}=(8\times10^{-12})^2 \mathrm{m}^2 \, \mathrm{Hz}^{-1} \, .
\end{equation}
Here, we neglect the impact of the confusion noise from the galactic foreground~\cite{Karnesis:2021tsh,Liu:2023qap}.
It is a non-stationary noise and does not have a significant impact on our results here.
In the limit of a large SNR, if the probability distribution of parameter measurement errors follows a Gaussian distribution~\cite{Cutler:1994ys,Poisson:1995ef,Vallisneri:2007ev}, the root-mean-square errors on the parameter $\theta_i$ can be obtained by
\begin{equation}
    \Delta\theta_i=\sqrt{(\Gamma^{-1})_{ii}} \, ,
\end{equation}
where the Fisher matrix
\begin{equation}
    \Gamma_{ij}=\left(\frac{\partial h}{\partial \theta_i}\bigg|\frac{\partial h}{\partial\theta_j}\right)\, .
\end{equation}
In our analysis, the parameters of the waveform are
\begin{equation}
    \theta=\left\{\mathcal{M},\eta,\chi,t_c,\phi_c,\beta\right\} \, ,
\end{equation}
where $\delta=(m_1-m_2)/M$, $\chi=(1+\delta)\chi_1/2+(1-\delta)\chi_2/2$ is the effective spin parameter of IMRPhenomB\cite{Ajith:2009bn}.
$t_c$ and $\phi_c$ represent the time and phase at the coalescence, respectively.
These quantities can be set to 0 in a specific coordinate system.
For the fiducial values of the ppE parameters, $\beta$ equals zero, as in the case of GR, since it encodes the effects arising from EdGB gravity.
The component masses of MBHBs $(m_1,m_2)=(2\times10^5M_\odot,10^5M_\odot)$, $(3\times10^5M_\odot,10^5M_\odot)$ and $(4\times10^5M_\odot,10^5M_\odot)$ with different spins configurations $(\chi_1,\chi_2)$ at redshifts $z=2,3,4,5$.

\section{\label{sec:results}Results}
The following are our results.
Assuming that GR is the correct theory, the standard deviation of each ppE parameter obtained from the Fisher matrix analysis shows the ability of Taiji to detect the specific modification introduced by EdGB.
In this paper, we include the leading $-1$PN order correction in the phase of the waveform and only consider the inspiral stage.
Considering that the phase corrections are derived within the context of the small coupling approximation.
As suggested in Ref.~\cite{Perkins:2021mhb}, for a binary system, the validity of the approximation can be assessed:
\begin{equation}
    16\pi\frac{\alpha^2_{EdGB}}{m^4_s}\lesssim\frac{1}{2} \, ,
\end{equation}
where $m_s$ is the smallest length scale of the system. 
In our analysis, we choose $m_2$ as the $m_s$ for MBHBs. 
Our results satisfy the small coupling approximation.

\begin{figure}
    \includegraphics[width=0.7\linewidth]{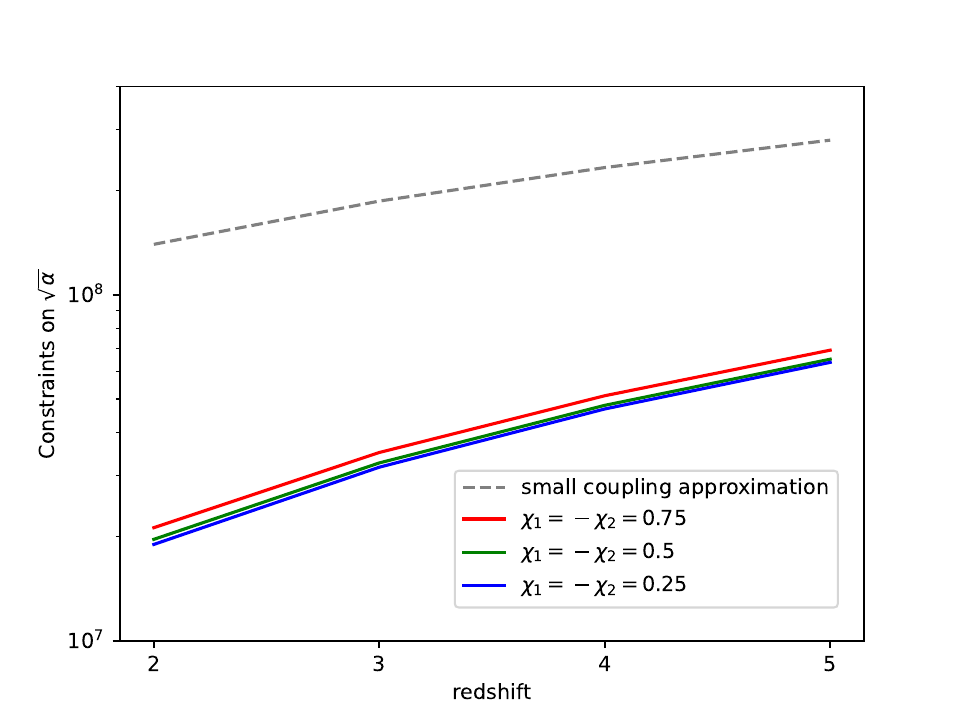}
    \caption{\label{fig:redshiftspin} Constraints on $\sqrt{\alpha}$ from MBHBs with spins $\chi_1=-\chi_2$ at different redshifts $z$. The dashed gray curve shows the small coupling limit.}
\end{figure}
Fig.~\ref{fig:redshiftspin} shows the constraints on $\sqrt{\alpha}$ given by the Fisher matrix and the small coupling limit.
For the MBHB with $(m_1,m_2)=(2\times10^5M_\odot,10^5M_\odot)$ at redshift $z=2$, the constraints on the coupling constant $\Delta\beta$ for the three different spins $(\chi_1,\chi_2)=(0.25,-0.25),(0.5,-0.5),(0.75,-0.75)$ are of the same order: $\Delta\beta\sim10^{-8}$, which corresponds to $\sqrt{\alpha}\sim10^7$m.
As the redshift increases, the uncertainty of the parameter $\sqrt{\alpha}$ also increases accordingly.

\begin{figure*}
    \includegraphics[width=0.49\linewidth]{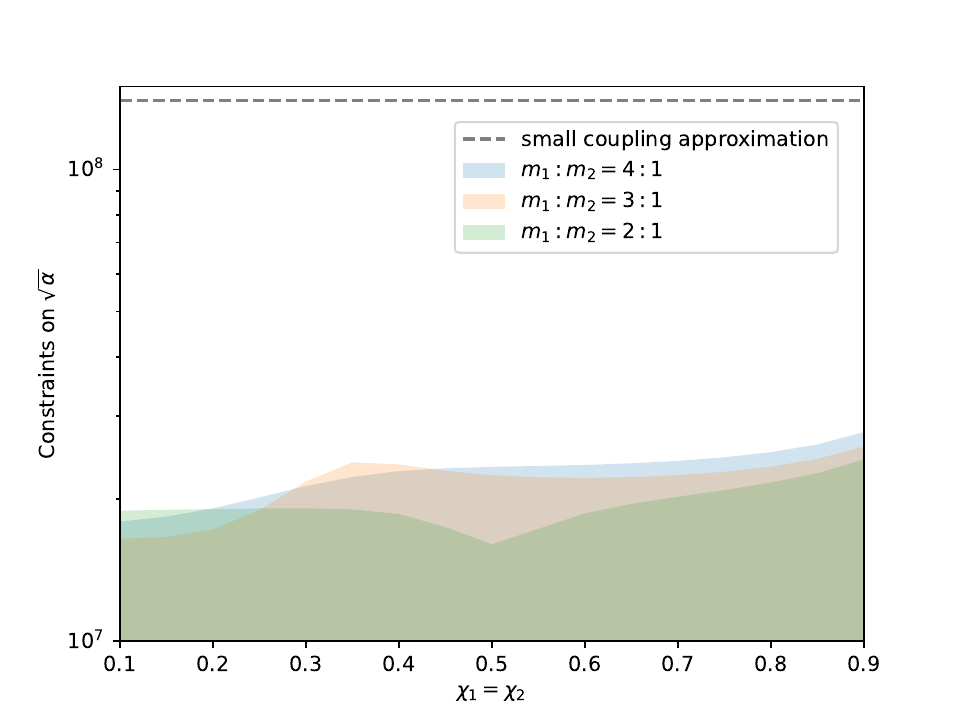}
    \includegraphics[width=0.49\linewidth]{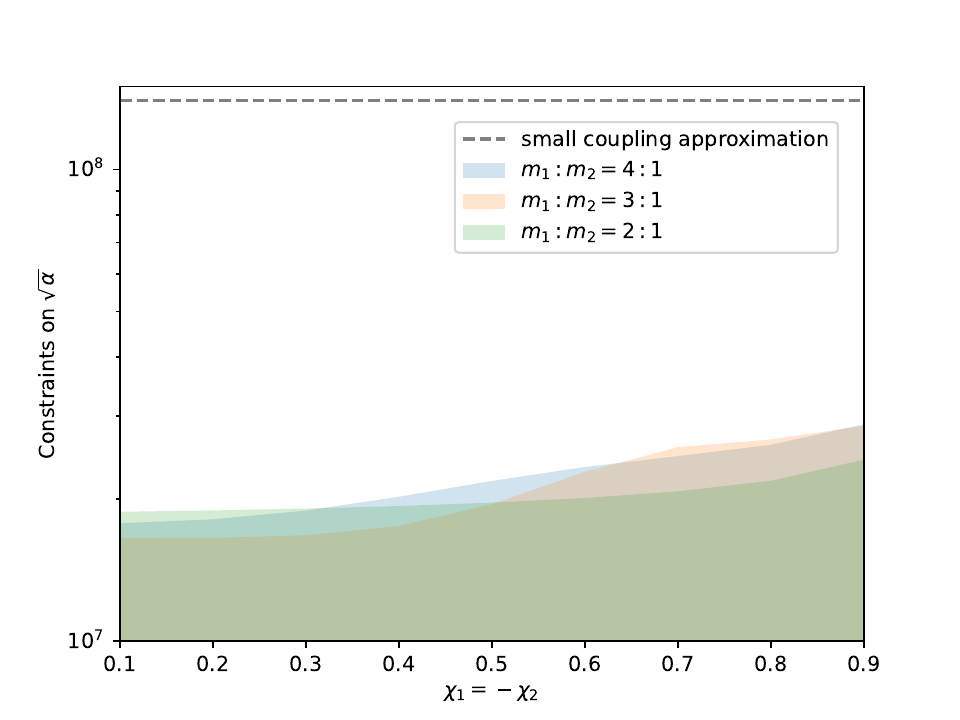}
    \caption{\label{fig:spinmassratio} The constraints on $\sqrt{\alpha}$ with the change in spins $\chi_1=\chi_2$ and $\chi_1=-\chi_2$ of the MBHBs with different mass ratios and $m_2=10^5M_\odot$ at redshift $z=2$. The dashed gray curve shows the small coupling limit.}  
\end{figure*}
Fig.~\ref{fig:spinmassratio} shows the constraints on $\sqrt{\alpha}$ with different mass ratios at redshift $z=2$.
In the case with $\chi_1=-\chi_2$, the measurement errors in $\sqrt{\alpha}$ tend to be larger as the spin increases.
But for $\chi_1=\chi_2$, the constraints on $\sqrt{\alpha}$ do not change monotonically with spin.
It is worth noting that the bounds obtained from the Fisher matrix analysis in this study satisfy the small coupling approximation, indicating the reliability of our results.

In previous studies~\cite{Yagi:2011xp,Cornish:2011ys}, the authors estimated the upper bound on the EdGB parameter $\Delta\beta$ with LISA, resulting in $\Delta\beta\lesssim10^{-6}$, which leads to $\sqrt{\alpha}\lesssim10^8m$.
Therefore, Taiji can put slightly tighter constraints on EdGB gravity compared to LISA.
In addition, comparable results were obtained by using the ringdown signals with space-based detectors~\cite{Shao:2023yjx}. 
The distinction is that the authors used a dimensionless parameter $\zeta=\frac{\alpha}{M^2}$ to characterize the effects from EdGB theory.
The accuracy of parameter estimation $\Delta\zeta$ spans from 0.015 to 0.5, which is similar to our results here.
With Tianqin, the constraints on $\sqrt{\alpha}$ will be less than $10^9m$
In Ref.~\cite{Shi:2022qno}, the authors investigated the constraints on ppE parameters at different PN orders with Tianqin.
Their findings exhibit the same order of magnitude as our results at $-1$PN order for MBHBs.
Our results here are also consistent with the case of multiband joint observations~\cite{Gnocchi:2019jzp}. 

In comparison to the constraints obtained from the ground-based detection, the results from LISA/Taiji are much weaker~\cite{Wang:2021jfc,Lyu:2022gdr,Wang:2023wgv}. 
This is because the parameter $\alpha$ is related to the mass or scale of the binary system. 
Consequently, the constraints on $\alpha$ will be tighter in the systems with smaller scales~\cite{Berti:2015itd,Wang:2023wgv}.

\section{\label{sec:sum}Summary}
In this paper we investigate the capability of Taiji to constrain the parameters of EdGB gravity by using the inspiral GW signals generated by MBHBs.
The signals are generated from spinning MBHBs with different spins and mass ratios at redshifts $z=2, 3, 4, 5$.
The observation time $T_{obs}=0.5 \mathrm{yr}$.
The Fisher matrix is employed to estimate the measurement errors of the phase parameters in the ppE waveform, including the $-1$PN order corrections arising from EdGB gravity.
Our results show that the  $1\sigma$ bounds on the parameter $\beta$ is $10^{-7}\sim10^{-8}$ and the corresponding constraints on $\sqrt{\alpha}$ is $\sim10^7m$.

To simplify the analysis, we employ the sky-averaged response function instead of the full detector response, neglecting the foreground noise from our Milky Way. 
Moreover, we do not include the effects from the Time Delay Interferometry~\cite{Tinto:2020fcc,Wang:2017aqq}, a crucial technique for noise suppression in space-based missions.
The potential enhancement from network observations~\cite{Cai:2023ywp} will also be left for future study.

\begin{acknowledgments}
This work is supported in part by the National Key Research and Development Program of China Grant No. 2020YFC2201501, in part by the National Natural Science Foundation of China under Grant No. 12147132, No. 12075297 and No. 12235019.
\end{acknowledgments}

\bibliographystyle{unsrt}
\bibliography{myref.bib}
\end{document}